\documentclass[preprint,showpacs,preprintnumbers,amsmath,amssymb]{revtex4}
\usepackage{CJK}

\usepackage{graphicx}
\usepackage{dcolumn}
\usepackage{bm}
\usepackage{epstopdf}

\newcommand{\feyn}[1]{#1\kern-0.45em/}

\begin{document}
\begin{CJK*}{UTF8}{gbsn}
	
\preprint{}

\begin{center}
{\Large\bf  Exclusive production of $B_c$ mesons in $e^+e^-$ colliders }\\[2cm]
{\large\bf  Yan-Bing Wei$^{a,b}$, Run-Hui Li$^{c}$,  Dan Luo$^{d}$, Cai-Dian L\" u$^{a,b}$ and Yue-Long Shen$^{d}$}\\[0.5cm]
 {\it $^a$Institute of High Energy Physics, CAS, P.O. Box 918(4),
Beijing 100049, P.R. China}\\
 {\it $^b$School of Physics, University of Chinese Academy of
Sciences, Beijing 100049, P.R. China}\\
{\it $^c$School of Physical Science and Technology, Inner Mongolia University, \\Huhhot 010021, P.R. China}\\
{\it $^d$ College of Information Science and Engineering, Ocean
University of China, Qingdao, Shandong 266100, P.R. China
 }\\[1.5cm]
\end{center}


\begin{abstract}
Within the framework of the perturbative QCD approach, we calculate the time-like
$B_cB_c(B_c^*)$ form factors $F(Q^2)$ and $A_2(Q^2)$. We include relativistic corrections and QCD corrections, either of which can give about $20\%$ correction to the leading-order
contribution, but there are cancellation effects between them. We
calculate the cross sections of the $e^+e^-\to B_c^-B_c^+(B_c^{*+})$
processes. The cross sections are enhanced at the $Z$ pole to be $\sigma^{PP}(Q=m_Z) \sim 1.3\times10^{-5}\text{pb}$ and
$\sigma^{PV}(Q=m_Z) \sim 2.5\times10^{-5}\text{pb}$, which
are still too small to be detected by proposed $e^+e^-$ colliders
 such as the Circular Electron Positron Collider.

\end{abstract}


\maketitle

\section { Introduction}

The study of the $B_c$ meson is of special interest, since it 
is the ground state of the doubly heavy-flavored ($\bar b c$) system, which
is unique in the Standard Model.  Both production and
decay processes of the $B_c$ meson contain rich heavy quark dynamics,
so they are worthy of systematic study. Experimentally,
observations of $B_c$ mesons are so far available only at hadron
colliders  \cite{cdf}. Hadron
colliders only provide limited knowledge of
the production of $B_c$ mesons because the total collision energy
cannot be well controlled. For $B_c$ production in
$e^+e^-$ colliders, the total collision energy
can be well controlled and many meaningful observables, such as
the angular distributions of final states, could be measured. Despite these advantages, the $B_c$ meson has
not yet been observed in $e^+e^-$ colliders \cite{lep-1}. The
production rate is too small to have been observed at LEP. Nowadays,
several high luminosity $e^+e^-$ facilities are proposed, such as
the International Linear Collider (ILC), Circular Electron Positron Collider (CEPC), and Future Circular Collider (FCC-ee). These $e^+e^-$ colliders could provide us with the
opportunity to produce $B_c$ mesons, especially at the $Z$-pole.

There are many studies of $B_c$ meson
decays \cite{Liu:2009qa,Liu:2010nz,Liu:2010da,Liu:2010kq,Xiao:2011zz,Rui:2015iia,Xiao:2013lia,Wang:2014yia,Liu:2017cwl,
Wang:2016qli,Yang:2010ba,Qiao:2012vt,Qiao:2012hp,Shen:2014msa,Wang:2015bka,Zhu:2017lqu}.
The semi-inclusive production of the $B_c$ meson has also been
investigated extensively \cite{changchen,Wu:2002ig,Chang:2005bf,Chang:2007si,Zheng:2017xgj}.
However, the knowledge of $B_c$ production in exclusive processes is
very limited. To deal with hard exclusive processes involving heavy
quark-anti-quark systems, two kinds of factorization approaches have
been proposed in the literature. One is the non-relativistic QCD
(NRQCD) factorization approach
\cite{Bodwin:1994jh,Brambilla:2004jw,Brambilla:2004wf,Brambilla:2010cs},
in which the amplitude of the process can be factorized into the
product of the
short-distance coefficient and NRQCD matrix-element. In the other
approach (called collinear factorization) the amplitude of the production
process can be expressed as a convolution of the
hard-kernel and the universal light-cone distribution amplitude
(LCDA) \cite{Lepage:1980fj,Chernyak:1983ej}. The LCDA of a large
boosted $B_c$ meson is defined by sandwiching the gauge invariant
non-local quark bilinear operators between the vacuum and the $B_c$
state. For the leading-twist LCDA, we have:
\begin{eqnarray}
\langle B_c(^1S_0,P)\vert (\bar b W_c)(\omega
n_+){n}\!\!\!\slash_+\gamma_5(W_c^\dag c)(0)(\omega)\vert 0\rangle
&=&- i f_{P}n_+P\int_0^1 dx ~e^{i\omega n_+Px} \hat{\phi}_{P}(x;\mu)\,, \\
\langle B_c(^3S_1,P,\varepsilon^*)\vert (\bar b W_c)(\omega
n_+){n}\!\!\!\slash_+(W_c^\dag c)(0)(\omega)\vert 0\rangle
&=& f_{V}m_V n_+\varepsilon^*\int_0^1dx~e^{i\omega n_+Px} \hat{\phi}_{V}^{\parallel}(x;\mu)\,,\\
\langle B_c(^3S_1,P,\varepsilon^*)\vert(\bar b W_c)(\omega
n_+){n}\!\!\!\slash_+\gamma^\alpha_\perp(W_c^\dag c)(0)(\omega)
\vert 0\rangle&=&f_{V}^{\perp}n_+P \varepsilon_\perp^{*\alpha}
\int_0^1 dx ~e^{i\omega n_+Px} \hat{\phi}_{V}^{\perp}(x;\mu)\,,
\end{eqnarray}\label{eq:def1}
where the Wilson-line
\begin{eqnarray}
W_c(x)=\textrm{P  exp} \Big[ ig_s\int_{-\infty}^0 ds~
n_+A(x+sn_{+}) \Big],
\end{eqnarray}
 and $\hat \phi_P(x)$, $\hat \phi_V(x)$ and $\hat
\phi_V^{\perp}(x)$ are LCDAs of pseudo-scalar, longitudinally
polarized and transversely polarized vector $B_c$ mesons,
respectively.  Using the ``re-factorization" scheme
\cite{Ma:2006hc}, the $B_c$-meson LCDA has been calculated to
next-to-leading order with respect both to the coupling constant
and the velocity of the heavy quark
\cite{Bell:2008er,Wang:2013ywc,Xu:2016dgp,Wang:2017bgv}.

Collinear factorization breaks down if there  exists an endpoint
singularity. A  solution is to introduce a transverse
momentum to the parton, which is called $k_T$ factorization. The
PQCD approach based on $k_T$ factorization has been employed to study a
large number of exclusive processes
\cite{Li:1992nu,Li:1994cka,Keum:2000ph,Keum:2000wi,Lu:2000em},
including pair production of light
mesons \cite{Lu:2007hr,Chen:2009sd,Hu:2012cp,Lu:2018obb}. In the
present paper we will employ the PQCD approach to study the
exclusive $B_c$ meson pair production processes, i.e. $e^+e^-\to
B_c^-B_c^+(B_c^{*+})$. The amplitude can be divided into the
leptonic part and the hadronic part. The hadronic part is actually the
time-like $B_cB_c$ or $B_cB_c^*$ form factors and evaluating these
form factors is the main object of this work. The form factors
will be calculated at the leading power in $1/Q$ (Q being the energy
of the $e^+e^-$ pair in the center-of-mass frame), while power
corrections from higher-twist LCDAs and quark mass effects will be
considered in the future.

This paper is organized as follows. In the next section we
calculate time-like $B_cB_c(B_c^*)$ form factors and cross
sections of $e^+e^-\to B_c^-B_c^+(B_c^{*+})$ processes using PQCD
approach. Numerical analysis is presented in the third section.
The last section gives a summary.

\section {Time-like form factors in the PQCD approach}
The differential cross section of $e^+(k')e^-(k)\to \gamma^*,Z^0 \to
B_c^-(P)B_c^+(P')(B_c^{*+})$ is
\begin{eqnarray}{d\sigma\over
d\Omega}={|{\bf k}|\over 32\pi^2 Q^3 }{1\over
4}\sum_{s,s'}|A_\gamma+A_Z|^2, \label{cross}\end{eqnarray} where
$s$ and $s'$ are the spins of the ingoing electron and positron respectively. The
momentum transfer $q=k+k'=P+P'$ and the energy scale
$Q=\sqrt{q^2}$. The amplitudes are
\begin{eqnarray}
A^{PP}_\gamma&=&\sum_{q=b,c}{-ie^2e_q\over Q^2 }\bar
v(k')\gamma_\mu u(k)(P
'-P)^\mu\, F^q(Q^2), \nonumber\\
A^{PP}_Z&=&{i\sqrt{2}G_Fm_W^2\over \cos^2
\theta_W}\sum_{q=b,c}{T_3^q-2e_q\sin^2 \theta_W\over
Q^2-m_Z^2+i\Gamma_Zm_Z }\bar v(k')\gamma_\mu(-{1\over 2}+2\sin^2
\theta_W +{1\over 2}\gamma_5) u(k)\nonumber \\ &\times
&(g^{\mu\nu}-{{q^\mu q^\nu}\over m_Z^2-i\Gamma_Zm_Z} )(P
'-P)_\nu\, F^q(Q^2),
\end{eqnarray}

\begin{eqnarray}
A^{PV}_\gamma&=&\sum_{q=b,c}{-ie^2e_q\over Q^2 }\bar
v(k')\gamma^\mu
u(k)\epsilon_{\mu\nu\rho\sigma}\varepsilon_{B_c^{*+}}^{*\nu}P^\rho
P^{'\sigma}\, V(Q^2),\nonumber\\
A^{PV}_Z&=&{i\sqrt{2}G_Fm_W^2\over \cos^2
\theta_W}\sum_{q=b,c}{1\over Q^2-m_Z^2+i\Gamma_Zm_Z }\bar
v(k')\gamma_\mu(-{1\over 2}+2\sin^2 \theta_W +{1\over 2}\gamma_5)
u(k)\nonumber\\&\times&(g^{\mu\lambda}-{{q^\mu q^\lambda}\over
m_Z^2-i\Gamma_Zm_Z} )\bigg \{[T_3^q-2e_q\sin^2
\theta_W]\epsilon_{\lambda\nu\rho\sigma}\varepsilon_{B_c^{*+}}^{*\nu}P^\rho
P^{'\sigma}\, V^q(Q^2)
\nonumber\\&-&T_3^q[A^q_1(Q^2)Q^2\varepsilon^{\ast\lambda}
\,+A^q_2(Q^2){\varepsilon^\ast\cdot q}(P '-P)^\lambda]\bigg\},
\end{eqnarray}
where $T_3^b=-1/2$ and $T_3^c=1/2$. The time-like  form factors
which appear in the above amplitudes are defined by:
\begin{eqnarray}
\!\!\left<B_c^+(P ')B_c^-(P)\left|\bar q\gamma_\mu
q\right|0\right>\!\!&=&(P '-P)_\mu\,
F^q(Q^2)\,,\nonumber \\
\!\!\left<B_c^{*+}(P ')B_c^-(P)\left|\bar q\gamma_\mu
q\right|0\right>\!\!&=&\epsilon_{\mu\nu\rho\sigma}\varepsilon_{B_c^{*+}}^{*\nu}P^\rho
P^{'\sigma}\, V^q(Q^2)\,,\nonumber \\
\!\!\left<B_c^{*+}(P ')B_c^-(P)\left|\bar q\gamma_\mu\gamma_5
q\right|0\right>\!\!&=& A^q_1(Q^2)(Q^2\varepsilon^{\ast}_\mu-{\varepsilon^\ast\cdot
q}\,q_\mu)
+\,A^q_2(Q^2){\varepsilon^\ast\cdot q}(P '-P)_\mu,
 \end{eqnarray}
 where the equation of motion has been employed.
Furthermore,
form factors $V^q(Q^2)$ and $A^q_1(Q^2)$ vanish at
leading power at tree level, and will not be considered in this work. Substituting the above amplitudes
into Eq.~(\ref{cross}), we have
\begin{eqnarray}
{d\sigma^{PP}\over d\Omega}&=&{1\over 64\pi^2 Q^2
}\bigg\{\{\sum_{q=b,c}[({e^2e_q\over Q^2}+Z_1^q(Q^2-m_Z^2)){\rm
Im}F_{q}+Z_1^qm_Z\Gamma_Z{\rm Re}F_{q}]\}^2\nonumber\\
&+& \{ \sum_{q=b,c}[({-e^2e_q\over Q^2}-Z_1^q(Q^2-m_Z^2)){\rm
Re}F_{q}+Z_1^qm_Z\Gamma_Z{\rm Im}F_{q}]\}^2\nonumber
\\&+&\{\sum_{q=b,c}[Z_2^q(Q^2-m_Z^2){\rm Im}F_{q}-Z_2^qm_Z\Gamma_Z{\rm
Re}F_{q}]\}^2\nonumber
\\&+&\{\sum_{q=b,c}[Z_2^q(Q^2-m_Z^2){\rm Re}F_{q}+Z_2^qm_Z\Gamma_Z{\rm
Im}F_{q}]\}^2\bigg\}{1\over 2}Q^4\sin^2 \theta,
\end{eqnarray}
where $Z_1^q={\sqrt{2}G_Fm_W^2\over \cos^2
\theta_W}{T_3^q-2e_q\sin^2 \theta_W\over
(Q^2-m_Z^2)^2+m_Z^2\Gamma_Z^2}(-{1\over 2}+2\sin^2 \theta_W)$ and
$Z_2^q={\sqrt{2}G_Fm_W^2\over \cos^2 \theta_W}{T_3^q-2e_q\sin^2
\theta_W\over (Q^2-m_Z^2)^2+m_Z^2\Gamma_Z^2}{1\over 2}$. For the
$B_cB_c^*$ channel, we have
\begin{eqnarray}
{d\sigma^{PV}\over d\Omega}&=&{1\over 32\pi^2}\bigg[{G_Fm_W^2\over
\cos^2 \theta_W}{1\over (Q^2-m_Z^2)^2+m_Z^2\Gamma_Z^2}\bigg]^2
\bigg[\{\sum_{q=b,c}[(Q^2-m_Z^2){\rm Im}A_{2}^q-m_Z\Gamma_Z{\rm
Re}A_{2}^q]T_3^q\}^2\nonumber
\\&+&\{\sum_{q=b,c}[(Q^2-m_Z^2){\rm Re}A_{2}^q+m_Z\Gamma_Z{\rm
Im}A_{2}^q]T_3^q\}^2\bigg]({1\over 2}+4\sin^4\theta_W-2\sin^2\theta_W)\nonumber \\
&\times&{Q^4\sin^2\theta\over 2} ({Q^2 \over 4m_{B_c^*}^2}-1).
\end{eqnarray}


In the proposed forthcoming accelerators, the collision  energy in the
center-of-mass frame is much larger than the masses of the  $b$ quark
and $c$ quark. In this paper, we concentrate on the leading
power result of the relevant time-like form factors. Power
suppressed contributions, such as quark mass effects and
corrections from higher-twist LCDA, which may be important when $Q^2$ is not
large enough, will be neglected in our calculations. In the PQCD
approach, form factors are factorized into a convolution of
the transverse momentum dependent (TMD) wave function and the hard
scattering kernel. The tree-level hard kernel can be obtained by
computing the Feynman diagrams plotted in Fig. \ref{fig: LO
diagrams }. In the calculation, we  choose the momentum
fractions of charm quarks in the $B_c^-$ meson and $B_c^{+(*)}$ meson to be $x$ and $y$
respectively. For form factors $F^q$, we have
\begin{eqnarray}
\label{eq:ff-hard2} F^{b}(Q^2)\!&=&\!\frac{2\pi f_{B_c}^2
C_F}{N_c}\, Q^2\,\int^1_0
dxdy\int^{\infty}_{0} b_1db_1 b_2 db_2\,\alpha_s(t)\,\left[-x\,{\mathcal P}_{PI}(x,b_1)\,{\mathcal P}_{PII}(y,b_2)\right]\nonumber\\
&&\times\,{\mathcal{H}}(x,y,Q,b_1,b_2)\,{\rm
exp}\left[-S(x,y,b_1,b_2,Q)\right]\,,\nonumber \\
F^{c}(Q^2)\!&=&\!\frac{2\pi f_{B_c}^2 C_F}{N_c}\, Q^2\,\int^1_0
dxdy\int^{\infty}_{0} b_1db_1 b_2 db_2\,\alpha_s(t)\,\left[- \bar x\,{\mathcal P}_{PI}(x,b_1)\,{\mathcal P}_{PII}(y,b_2)\right]\nonumber\\
&&\times\,{\mathcal{H}}(\bar x,\bar y,Q,b_1,b_2)\,{\rm
exp}\left[-S(x,y,b_1,b_2,Q)\right]\,,
\end{eqnarray}
where $\bar x=1-x$. The hard function is:
\begin{eqnarray}
\label{eq:kernel_t}
{\mathcal H}(x,y,Q,b_1,b_2)\!\!&=&\!\!\left(\frac{i\pi}{2}\right)^2H^{(1)}_0(\sqrt{xy}\,Qb_2)
\left[\theta(b_1-b_2)H^{(1)}_0(\sqrt{x}\,Qb_1)J_0(\sqrt{x}\,Qb_2)\right.\nonumber\\
&&+\left.
\theta(b_2-b_1)H^{(1)}_0(\sqrt{x}\,Qb_2)J_0(\sqrt{x}\,Qb_1)\right]\,.
\end{eqnarray}
The TMD wave function is:
\begin{eqnarray}
{\cal P}_P(x,\mathbf{b})&=&\hat\phi_P(x) \Sigma(x,{\bf b}),
\end{eqnarray}
where $\Sigma(x,b)$ is the transverse momentum dependent part. The LCDAs
can be expressed in the following expansions,
\begin{subequations}
\begin{eqnarray}
 \hat \phi_P(x)&=&   \hat \phi_P^{(0,0)}(x)+  \hat \phi_P^{(1,0)}(x)+  \hat \phi_P^{(0,1)}(x), \\
 \hat \phi_V^{||}(x)&=&   \hat \phi_V^{||(0,0)}(x)+  \hat \phi_V^{||(1,0)}(x)+ \hat \phi_V^{||(0,1)}(x),
\end{eqnarray}
\end{subequations}
where the superscript $(i,j)$ 
denotes the order of $\alpha_s$ and $v^2$-expansion. In order to
express the form factors in terms of $\alpha_s$ expansion, we
write
\begin{subequations}
\begin{eqnarray}
{\cal P}(x,\mathbf{b})&=&{\cal P}^{(0)}(x,\mathbf{b})+{\cal
P}^{(1)}(x,\mathbf{b})+...,\\
 {\cal P}^{(0)}(x,\mathbf{b})&=&   [\hat \phi^{(0,0)}(x)+  \hat \phi^{(0,1)}(x)] \Sigma(x,b), \nonumber \\
 {\cal P}^{(1)}(x,\mathbf{b})&=&   \hat \phi^{(1,0)}(x)\Sigma(x,b).
\end{eqnarray}
\end{subequations}

A typical value of $x$ is $x\sim x_0={m_c\over m_b+m_c}$. We have
$x\,Q^2\gg k_T^2$ when the energy scale $Q$ is large, thus we can
drop the transverse momentum in the quark propagator. Then form
factors can then be simplified as
\begin{eqnarray}
F^b_{(LO)}(Q^2) &=& i\frac{\pi^2 f_{B_c}^2 C_F\alpha_s(\mu) }{N_c}
\int_0^1dx dy \int_0^\infty db b \,{\mathcal P}_{PI}(x) {\mathcal
P}_{PII}(y) \nonumber
\\
&\times& \exp[-S_{\text{I}} (x,y, b, Q, \mu)] H_0^{(1)}(\sqrt{x
y}Qb),\\
 F^c_{(LO)}(Q^2)
&=& i\frac{\pi^2 f_{B_c}^2 C_F\alpha_s(\mu) }{N_c} \int_0^1dx dy
\int_0^\infty db b \,{\mathcal P}^{(0)}_{PI}(x) {\mathcal
P}^{(0)}_{PII}(y) \nonumber
\\
&\times& \exp[-S_{\text{I}} (x,y, b, Q, \mu)]
H_0^{(1)}\left(\sqrt{\bar{x}\bar{y}}Qb\right),
\end{eqnarray}
where {$S_{\text{I}} (x, y, b, Q,\mu) \equiv S (x, b, y, b, Q,
\mu) $} and the Hankel function {$H^{(1)}_0(x)=J_0(x)+iY_0(x)$}.
Form factors $A^{(b,c)}_2$ are related to $F^{(b,c)}$ at tree
level, as shown in Eqs. (\ref{eq.app1},\ref{eq.app2}):
\begin{eqnarray}
A^b_{2,(LO)}(Q^2) &=& -\frac{\pi^2 f_{B_c}f^\parallel_Vm_V
C_F\alpha_s(\mu) }{Q^2N_c} \int_0^1dx dy \int_0^\infty db b
\,{\mathcal P}_{PI}(x) {\mathcal P}_{V^{||}II}(y) \nonumber
\\
&\times& \exp[-S_{\text{I}} (x,y, b, Q, \mu)] H_0^{(1)}(\sqrt{x
y}Qb), \nonumber
\\
A^c_{2,(LO)}(Q^2) &=& \frac{\pi^2 f_{B_c}f^\parallel_Vm_V
C_F\alpha_s(\mu) }{Q^2N_c} \int_0^1dx dy \int_0^\infty db b
\,{\mathcal P}_{PI}(x) {\mathcal P}_{V^{||}II}(y) \nonumber
\\
&\times& \exp[-S_{\text{I}} (x,y, b, Q, \mu)]
H_0^{(1)}\left(\sqrt{\bar{x}\bar{y}}Qb\right).
\end{eqnarray}

\begin{figure}[h]
\begin{center}
\includegraphics[width=0.6  \columnwidth]{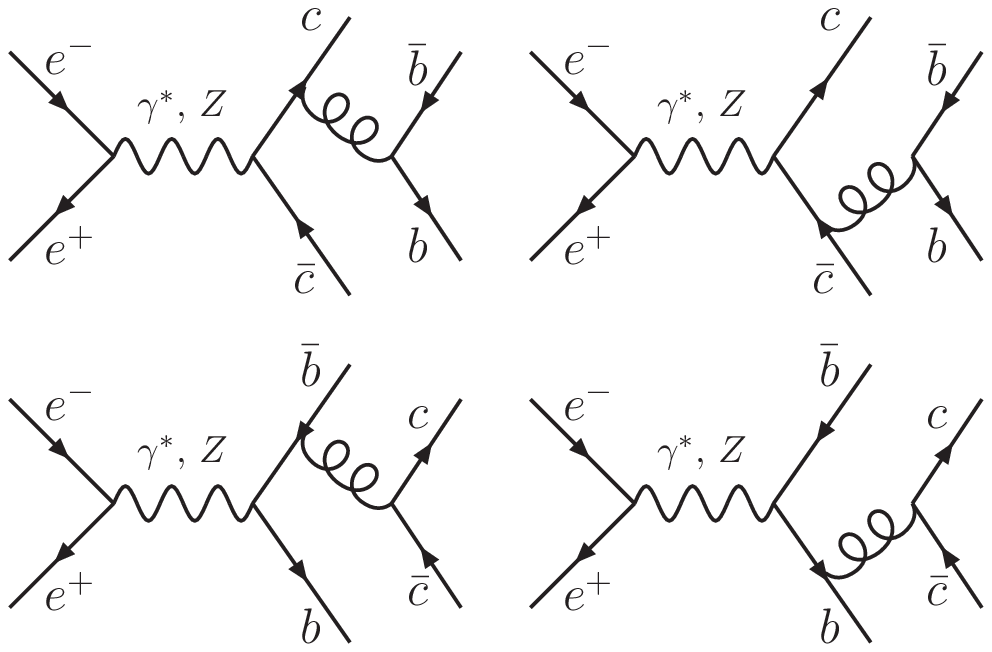}\\
 \caption{Feynman diagrams for $e^+e^-\to B^-_c B^{+(*)}_c$.} \label{fig: LO diagrams }
\end{center}
\end{figure}

The QCD corrections to form factors are of great importance in
theoretical analysis. In the PQCD approach,
NLO corrections to the time-like pion electromagnetic form
factors have been
 calculated by Li et al \cite{Hu:2012cp}.  In this paper, we consider
 NLO corrections to the form factors $F^q$. The calculation of NLO corrections in the PQCD approach
 is tediously complicated. To simplify the PQCD calculation, the hierarchy
$xQ^2, \,yQ^2\gg xyQ^2\sim k_{T}^2$ is postulated. As the quark
masses have
been neglected, the NLO hard kernel can be borrowed directly from
\cite{Hu:2012cp}:
\begin{eqnarray}
H^{(\text{NLO})}_{\text{F}}(x,k_{1T},y,k_{2T},Q^2,\mu) =
h(x, y, \delta_{12},Q,\mu)
H^{(\text{LO})}_{\text{F}}(x,k_{1T},y,k_{2T},Q^2) , \label{EMH1}
\end{eqnarray}
where
\begin{eqnarray}
h(x, y, \delta_{12},Q,\mu) &=& { \alpha_s(\mu) C_F \over 4 \pi }
\bigg[ -{3 \over 4}\ln { \mu^2 \over Q^2} - {17 \over 4}\ln^2 x  +
{27 \over 8} \ln x \ln y - {13 \over 8} \ln x + {31 \over 16} \ln
y
\nonumber
\\
& -&\!\ln^2 \delta_{12} +\!\left({17 \over 4} \ln x + {23 \over 8}
+i2\pi \right) \ln \delta_{12} +{  \pi^2 \over 12} + {1 \over 2}
\ln 2 + {53 \over 4}-i{3\pi\over 4}
\bigg] , \label{h_1}
\end{eqnarray}
and
\begin{eqnarray}
\ln{\delta_{12}} \equiv \ln{\frac{\left| {|{\bf k}_{1T}+{\bf
k}_{2T}|}^2 - x y Q^2 \right|}{Q^2}} +i\pi\Theta\left({|{\bf
k}_{1T}+{\bf k}_{2T}|}^2 - x y Q^2 \right) . \label{DEL12}
\end{eqnarray}

Including the NLO hard kernel and performing a Fourier transform of
Eq.~(\ref{EMH1}), we derive the TMD factorization formula of form
factors $F^q$ at leading power,
\begin{eqnarray}
F^b_{NLO}(Q^2) =  && i\frac{\pi f_{B_c}^2 C_F^2}{4N_c} \int_0^1dx dy
\int_0^\infty db b \, \alpha_s^2(\mu) {\mathcal P}^{(0)}_{PI}(x)
{\mathcal P}^{(0)}_{PII}(y) \exp[-S_{\text{I}} (x, y, b, Q, \mu)]
\nonumber \\
&&\times \left[
    \,\widetilde{h}(x, y, b, Q, \mu) \, H_0^{(1)}(\sqrt{x y}Qb)
+H^{(1)\prime \prime}_0\left(\sqrt{x y}Qb\right) \right]
,\nonumber  \\
F^c_{NLO}(Q^2) =  && i\frac{\pi f_{B_c}^2 C_F^2}{4N_c} \int_0^1dx dy
\int_0^\infty db b \, \alpha_s^2(\mu) {\mathcal P}^{(0)}_{PI}(x)
{\mathcal P}^{(0)}_{PII}(y) \exp[-S_{\text{I}} (x, y, b, Q, \mu)]
\nonumber \\
&&\times \left[
    \,\widetilde{h}(\bar x, y, b, Q, \mu) \, H_0^{(1)}(\sqrt{\bar x \bar y}Qb)
+H^{(1)\prime \prime}_0\left(\sqrt{\bar x \bar y}Qb\right)
\right],\label{EM1I1}
\end{eqnarray}
where
\begin{eqnarray}
\widetilde{h}(x, y, b, Q, \mu) &= & -{3 \over 4}\ln { \mu^2 \over
Q^2} -{1\over 4}\ln^2 \frac{4 x y}{Q^2 b^2} +\left( {17\over 8}\ln
x +{23\over 16}+\gamma_E+i{\pi \over 2} \right) \ln \frac{4 x
y}{Q^2 b^2}
\nonumber\\
& -& {17 \over 4}\ln^2 x  + {27 \over 8} \ln x \ln y - \left( {13
\over 8} +{17\gamma_E \over 4} -i{17 \pi\over 8} \right)\ln x +
{31 \over 16} \ln y
\nonumber\\
& -& {\pi^2 \over 2} + (1-2\gamma_E)\pi +{1 \over 2}\ln 2+ {53
\over 4} - {23 \over 8}\gamma_E - \gamma_E^2 + i\left({171\over
16}+\gamma_E\right)\pi . \label{th_1}
\end{eqnarray}

\section{Numerical analysis}

The most important nonperturbative input is  the LCDA of the $B_c$ meson,
which can be studied with non-relativistic QCD effective theory as both
internal quarks are heavy. At tree level and at leading order in
the expansion over the relative velocity, the quark and the
antiquark simply share the
momentum of the meson according to their masses,
  \begin{eqnarray}
  \hat\phi_P^{(0,0)}(x)&=&\hat\phi_V^{\parallel(0,0)}(x)=\delta(x-x_0).\,
  \end{eqnarray}
  If we include the gluon exchange effect and the power correction relevant
 to the quark velocity, the  parton configuration will be
 changed.  Explicit expressions of $\hat{\phi}^{(0,1)}(x)$
are given by~\cite{Xu:2016dgp}
  \begin{eqnarray}
  \hat\phi_P^{(0,1)}(x)&=&-\frac{ \langle \textbf{q}^2\rangle_P}{M^2}\Big[ \frac{2(1-2x_0)}{3x_0\bar x_0}\delta'(x-x_0)-\frac{1}{6}\delta''(x-x_0) \Big]\,,\\
  \hat\phi_V^{\parallel (0,1)}(x)&=&\hat\phi_V^{\perp(0,1)}(x)=-\frac{ \langle \textbf{q}^2\rangle_V}{M^2}\Big[ \frac{2(1-2x_0)}{3x_0\bar x_0}\delta'(x-x_0)-\frac{1}{6}\delta''(x-x_0) \Big]\,,
  \end{eqnarray}
where $M=m_b+m_c$ and $\langle \textbf{q}^2\rangle_{P,V}$ are the mean values of
$\textbf{q}^2$ in scalar and vector $B_c$ mesons respectively. For
QCD corrections to the LCDAs, we have
\begin{eqnarray}
 \hat \phi_P^{(1,0)}(x;\mu)&=&\frac{\alpha_s}{4\pi}C_F \bigg\{\Phi_1(x,x_0)
\bigg\} \,, \\
   \hat \phi_V^{\parallel(1,0)}(x;\mu)&=&\frac{\alpha_s}{4\pi}C_F \bigg\{\Phi_1(x,x_0)
-4\left[\frac{x}{x_0}\theta(x_0-x)+\left(  x\leftrightarrow \bar
x, x_0\leftrightarrow \bar x_0   \right)\right]_{+}\bigg\} \,,
\label{eq:INNRQCD}
\end{eqnarray}

with
\begin{eqnarray}
\Phi_1(x,x_0)&=&2\left[ \left( \ln{\frac{\mu^2}{M^2(x_0-x)^2}}-1 \right)\left( \frac{x_0+\bar x}{x_0-x}\frac{x}{x_0}\theta(x_0-x)+\left( x\leftrightarrow \bar x,  x_0\leftrightarrow \bar x_0   \right) \right) \right]_+ \nonumber\\
&&~~+\left[ \frac{4x\bar x}{(x_0-x)^2}\right]_{++}+\left[ 4x_0\bar
x_0 \ln{\frac{x_0}{\bar x_0}}+2(2x_0-1)
\right]\delta^\prime(x-x_0)\,.
\end{eqnarray}
Here we adopt the NDR scheme of $\gamma_5$, as the NLO hard
kernel is also obtained in this scheme.  The $++$- and
$+$-distributions are defined as
\begin{subequations}
\begin{eqnarray}
\int _0^1dx\Big[ f(x)  \Big]_{++}g(x)&=&\int _0^1dx f(x)(g(x)-g(x_0)-g'(x_0)(x-x_0)),  \\
\int _0^1dx\Big[ f(x)  \Big]_{+}g(x)&=&\int _0^1dx
f(x)(g(x)-g(x_0))\,,
\end{eqnarray}
\end{subequations}
where $g(x)$ is a smooth test function. Because LCDA with $v$ and
$\alpha_s$ corrections contains a Dirac-$\delta$ function or plus
distribution, we need to integrate over the momentum fraction $x$
first. The results  are given in the Appendix.

For the TMD wave function, we use
\begin{eqnarray}
\Sigma(x,\mathbf{b})&=&\exp\Big(-\frac{b^2}{4\beta^2}
\Big).
\end{eqnarray}
The  input parameters are listed in Table \ref{Parameters used
for numerical analysis.}.

\begin{table}[!h]
\centering
\begin{tabular}{|c|c|c|c|}
\hline
parameter & value & parameter & value \\
\hline
$m_b$ & 4.8 GeV & $m_c$ & 1.6 GeV \\
\hline
$m_W$ & 80.425 GeV & $m_Z$ & 91.1876 GeV\\
\hline
$m_{B_c}$ & 6.2749 & $\langle \textbf{q}^2\rangle$& 1.59 GeV$^2$\\
\hline
$\Gamma_Z$& 2.4952 GeV & $\text{sin}^2\theta_W$ & 0.23129  \\
\hline
$n_f$ & 5 & $\Lambda$ & 0.217 GeV \\
\hline
$\gamma_E$ & 0.57721566 & $G_F$ & 1.166391$\times10^{-5}$  GeV$^{-2}$\\
\hline
$\beta$ & 2 GeV$^{-1}$ & $f_{B_c}$ & 0.489$\pm$0.005 GeV \\
\hline
\end{tabular}
\caption{Parameters used for numerical analysis.}\label{Parameters
used for numerical analysis.}
\end{table}


\begin{figure}[!h]
\begin{center}
\includegraphics[height=6.5cm ]{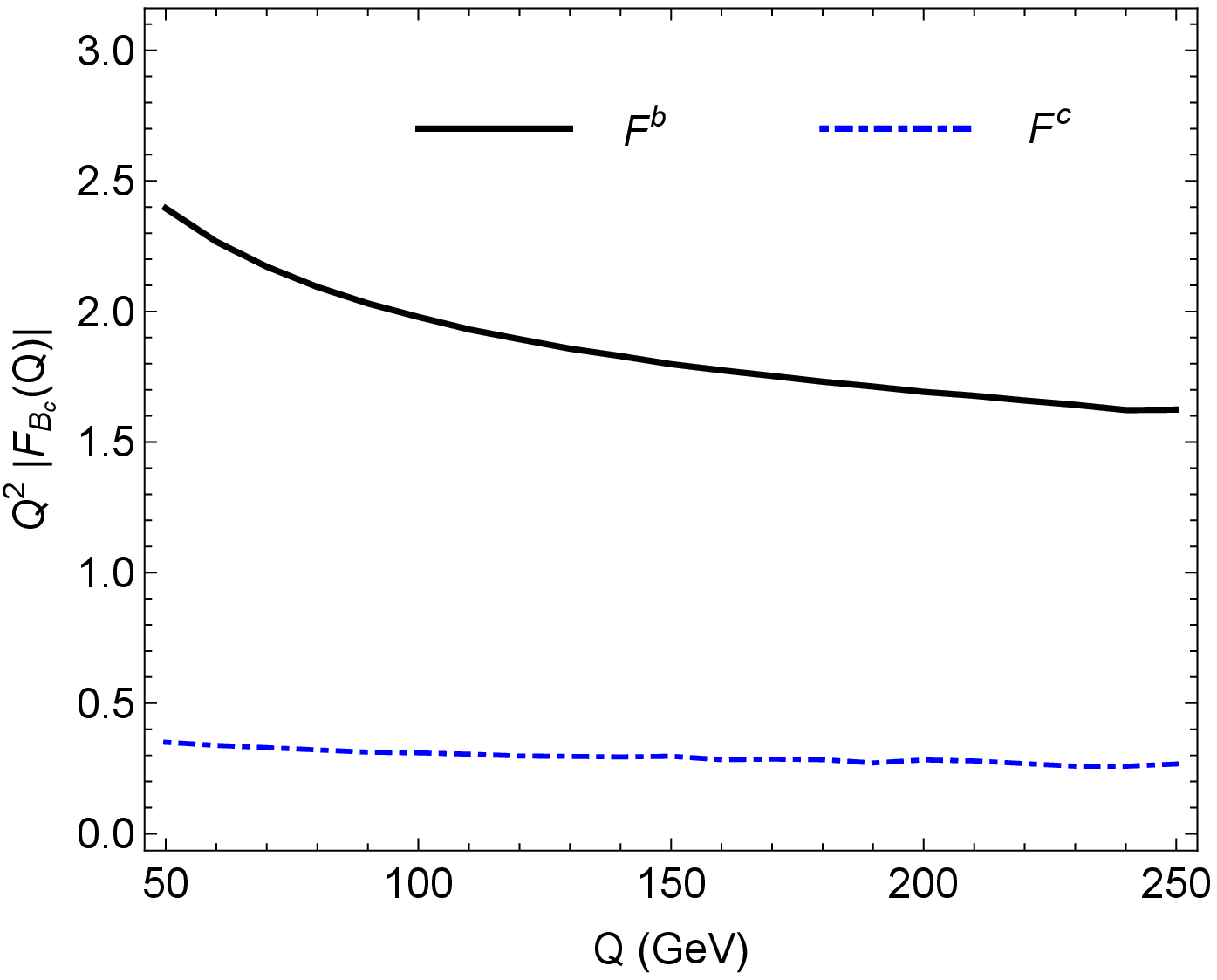}
\includegraphics[height=6.5cm ]{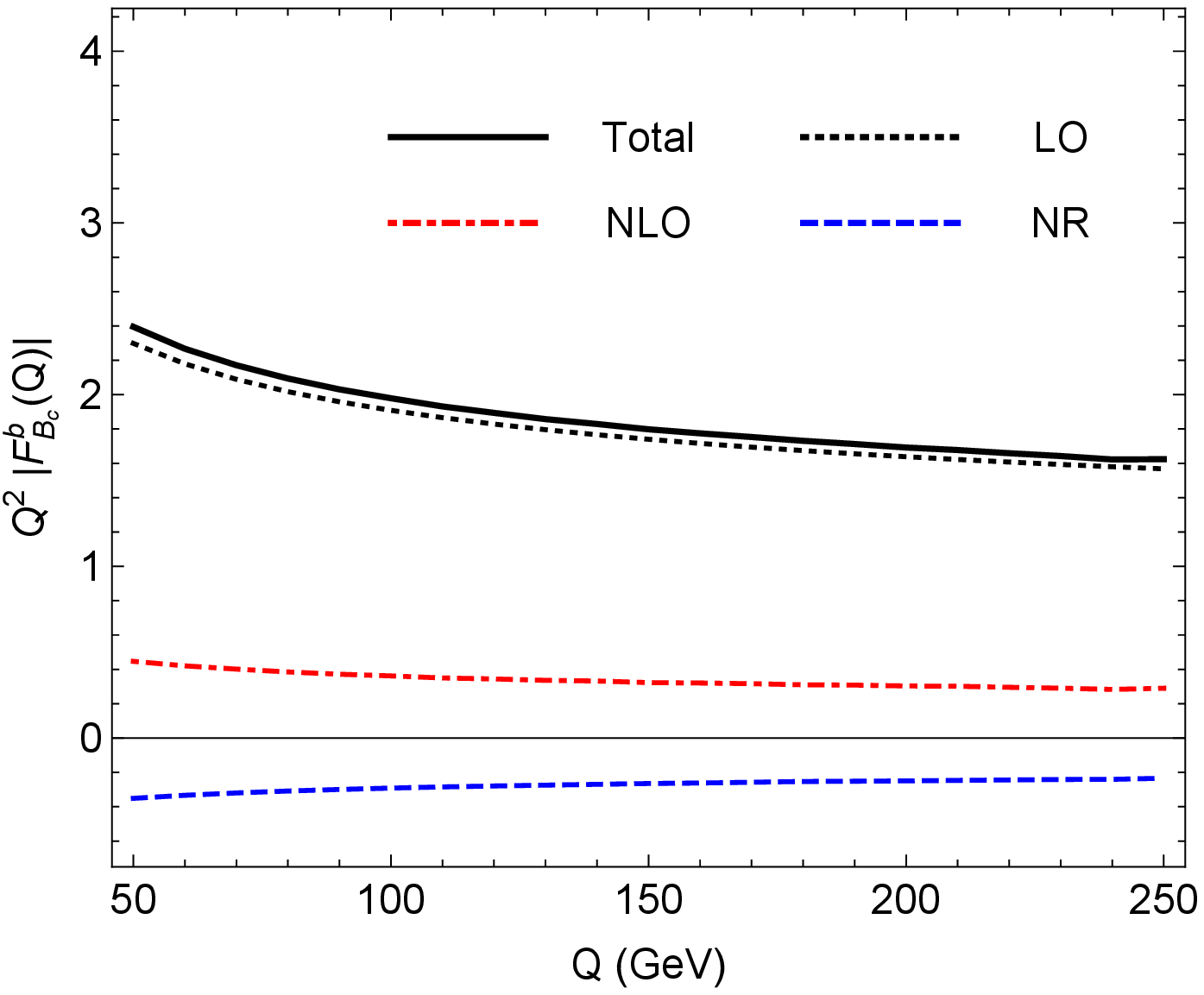}
\\ \hspace{0.5cm}(b)\hspace{-8cm}(a)
 \caption{Shape of $Q^2\vert F^c_{B_c}\vert $ and $Q^2\vert F^b_{B_c}\vert $. Left: the total results of $F^b$ and $F^c$ are presented by the solid black  curve and the  dot-dashed blue curve respectively. Right: the solid black, dotted black,  dot-dashed red and  dashed blue line are the total result, leading order results, NLO correction and relativistic correction of $F^b$, respectively.} \label{fig: fbc}
\end{center}
\end{figure}

We  are now ready to evaluate the numerical results of the form
factors and cross sections. We first concentrate on the $Q$
dependence of the time-like form factors. In the PQCD approach the
time-like form factors are complex, as the internal quark line may
be on-shell. The absolute values of form factors $F^c$ and $F^b$
are plotted in Fig. \ref{fig: fbc}. We do not show form
factors $A^c_2$ and $A^b_2$, since they are related to $F^c$ and
$F^b$. From the left-hand plot, we can see that the absolute value of form factor $F^c$ (the  dot-dashed blue curve) is
about an order or magnitude smaller than that of $F^b$ (the solid black  curve) because the wave
function of $B_c$ meson is not symmetrical between charm quark and
bottom quark, so the invariant mass of the internal gluon
propagator is much larger in $F^c$. To illustrate the effects of
relativity and the QCD corrections we
plot the form factor $F^b$ with these contributions. As we can see from the right-hand plot,
contributions from both relativistic corrections (the dashed blue  curve) and NLO corrections (the  dot-dashed red curve) are  about $20\%$ of the LO contribution (the dotted black  curve). However, there is
cancellation between the two kinds of corrections and the total
result is very close to the leading order contribution. Power
corrections from higher-twist LCDAs and quark masses are not
included, but they are not very significant at large
$Q^2$.

\begin{figure}[!h]
\begin{center}
\includegraphics[width=0.46  \columnwidth]{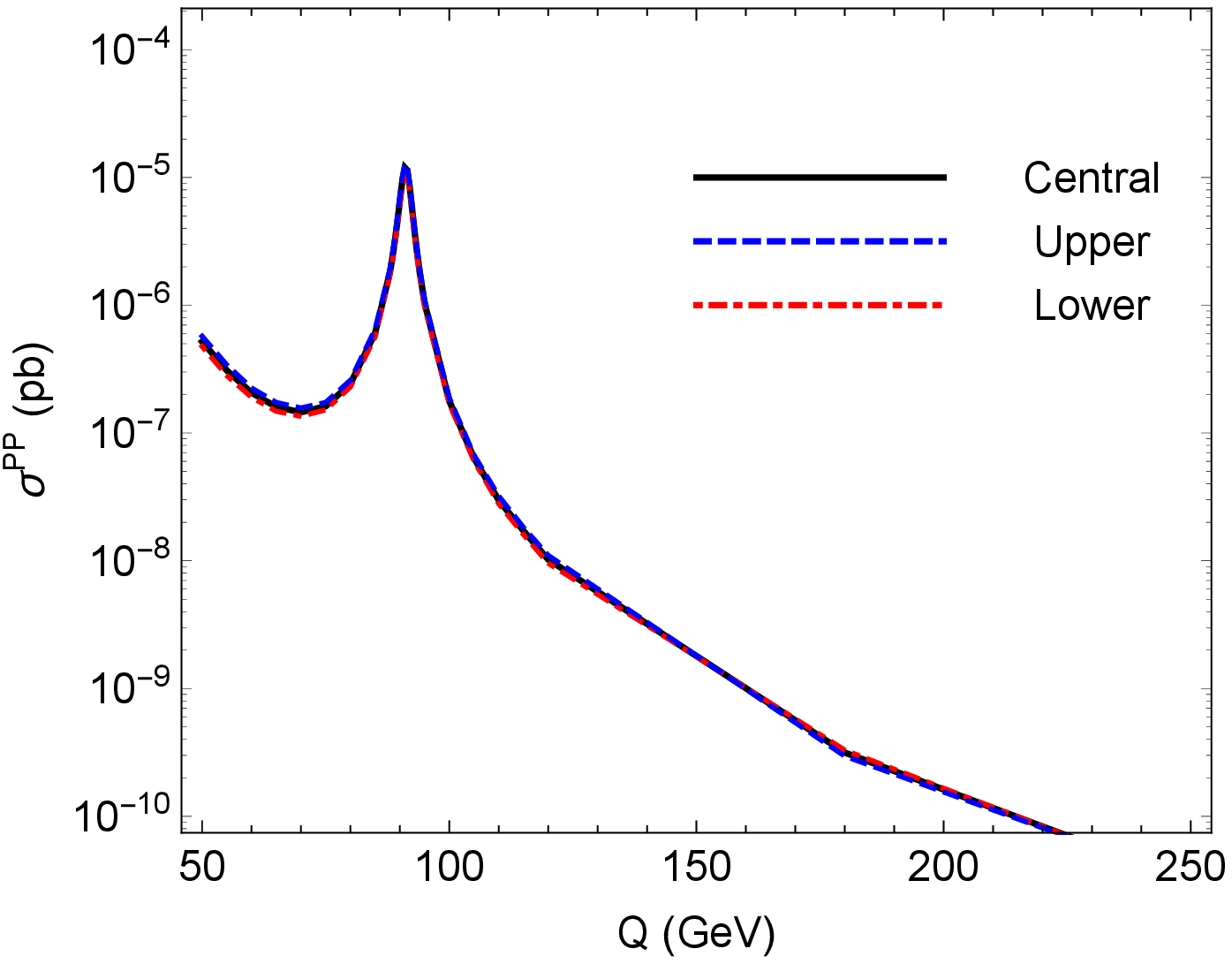}
\includegraphics[width=0.44  \columnwidth]{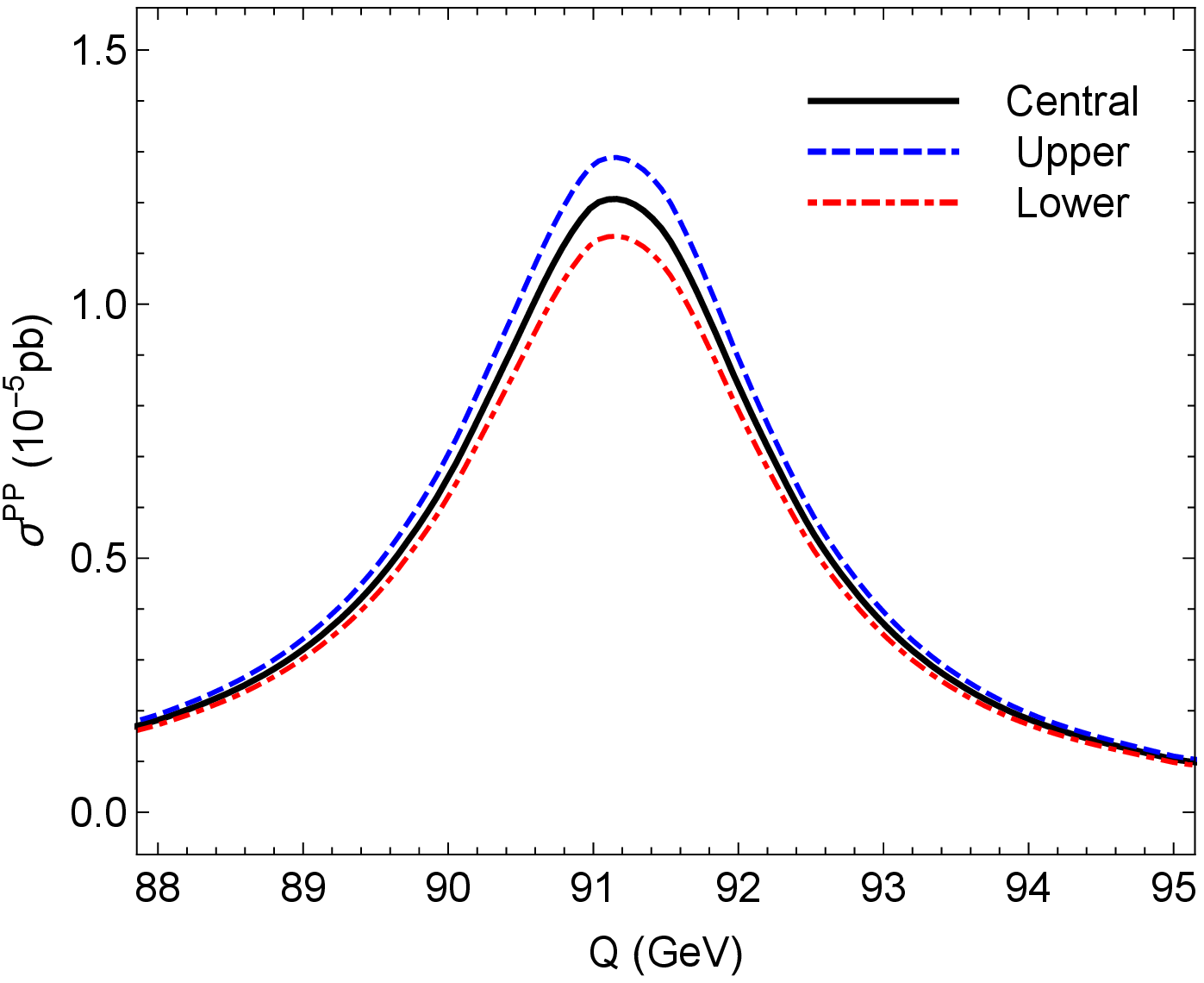}\\
\includegraphics[width=0.46  \columnwidth]{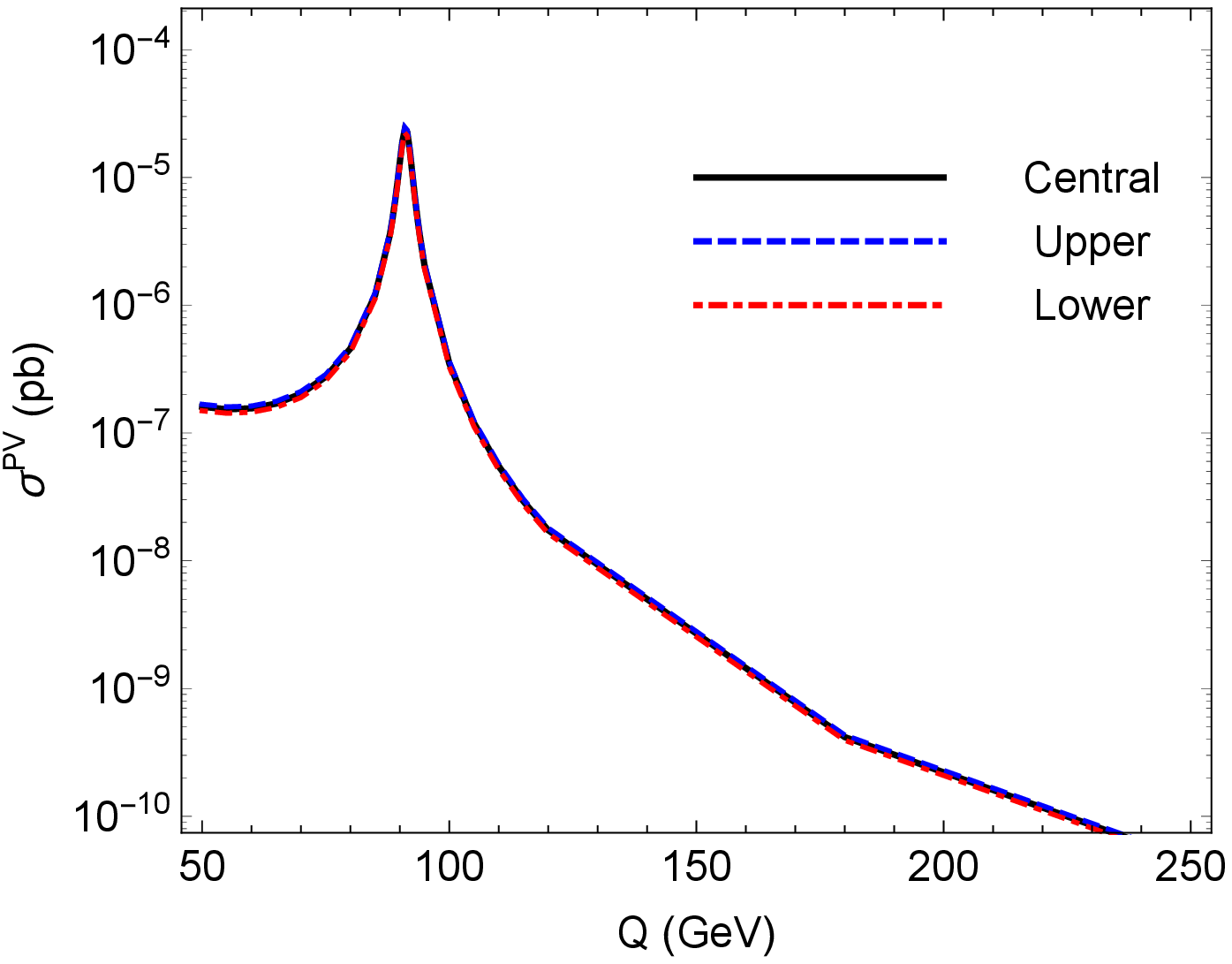}
\includegraphics[width=0.44  \columnwidth]{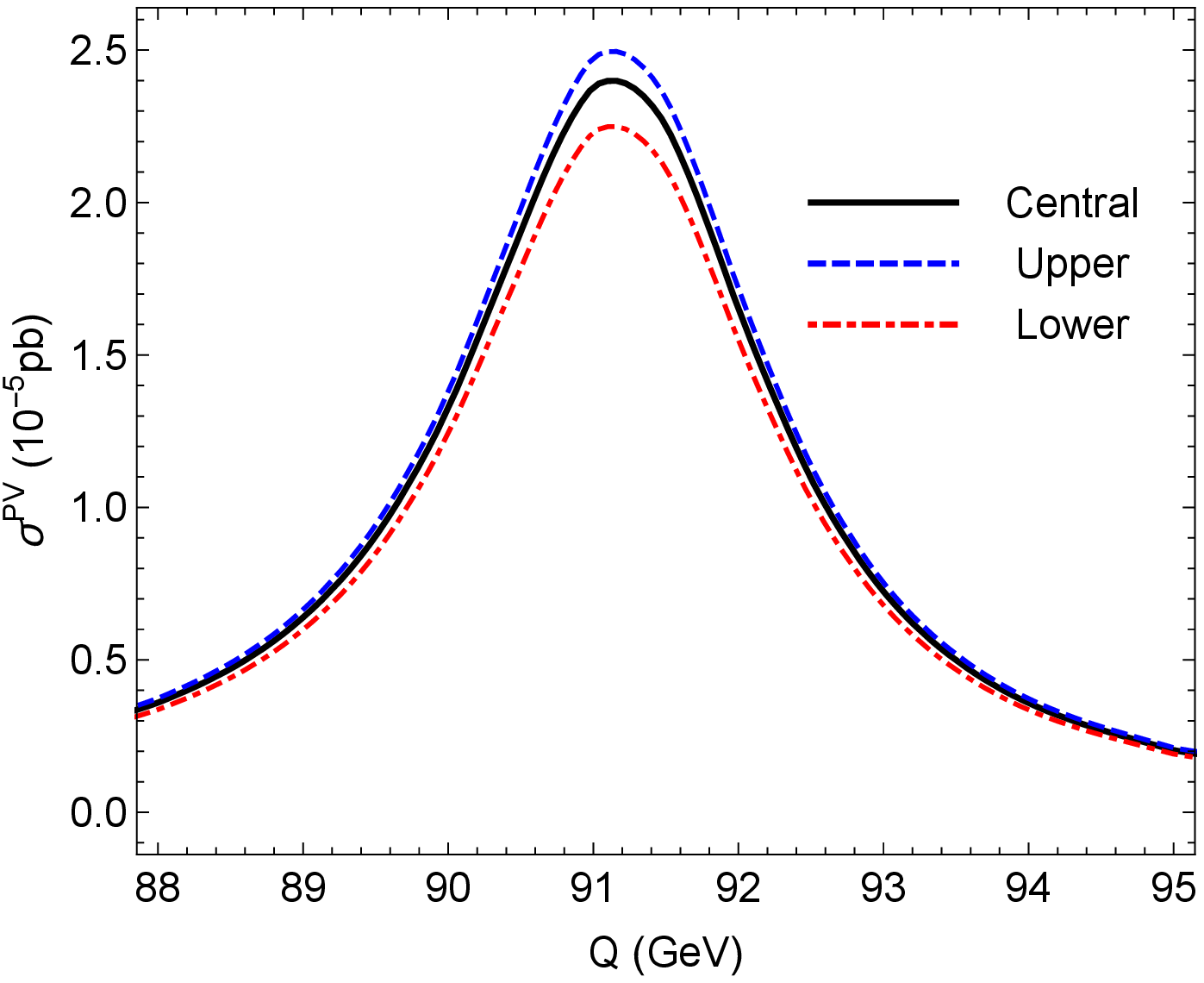}
 \caption{Cross sections of $e^+e^- \to B^-_cB^{+(*)}_c$ with errors from variations of the factorization scale $\mu$ and $B_c$-meson decay constant $f_{B_c}$. The solid black, dashed blue  and  dot-dashed red 
 curves correspond to central, upper and lower values of 
 cross sections, respectively. The kinematic region of the left two diagrams 
 is $50 - 250$ GeV and that of the right two diagrams is around the Z pole.} \label{fig: ppv}
\end{center}
\end{figure}

Taking advantage of the form factors computed with the PQCD approach,
the total cross sections of $e^+e^- \to B^-_cB^{+(*)}_c$ are
depicted in Fig. \ref{fig: ppv}. The factorization scale $\mu$
is taken to be $\mu=Q$ for the central value and $\mu\sim [Q/2,\,2Q]$
for the error estimate. The solid black  curves correspond to central values
and the  dashed blue and  dot-dashed red lines represent the errors of the cross sections. Considering $f_{B_c}$ only causes tiny errors because of its high accuracy, the cross sections at our kinematic region are not sensitive to the variation of factorization scale. As
expected, the cross section has a significant peak at the $Z$ pole
($Q=m_Z$) and the maximum values of the cross sections are
$\sigma^{PP}(Q=m_Z) \sim 1.3\times10^{-5}\text{pb}$ and
$\sigma^{PV}(Q=m_Z) \sim 2.5\times10^{-5}\text{pb}$. They are so
small that it is almost hopeless to detect  a $B^-_cB^{+(*)}_c$
pair in an $e^+e^-$ collider at the Z pole. This result is quite
different from $B\bar B$ production, which is enhanced by a low
energy resonance. For single $B_c$ production in semi-inclusive
processes, the cross section is much larger \cite{Zheng:2017xgj},
and can be reached by a high luminosity $e^+e^-$ collider.


\section{Summary}
$B_c$ production processes are important because the $B_c$ meson has
unique properties and is worthy of a thorough study. Within the
framework of the PQCD approach, we calculated the time-like form
factors $F^q(Q^2)$ and $A^q_2(Q^2)$ at  leading power in $1/Q$. The form
factors are factorized into a convolution of the transverse
momentum dependent wave function and the hard kernel. The wave
functions employed in this work, including QCD corrections and relativistic corrections,
 have recently been studied using NRQCD.
We evaluate the effects of relativistic corrections and QCD
corrections of the form factors numerically. Both relativistic and QCD
corrections can give about $20\%$
correction to the LO contribution, but there are cancellation
effects between them. We further worked out the cross sections of
$e^+e^-\to B_c^-B_c^+(B_c^{*+})$ processes. The cross sections are
enhanced at the $Z$ pole, but they are still too small to be
detected in future accelerators such as the CEPC.

\section*{Acknowledgement}
We are grateful to Q. A. Zhang for useful discussions and comments.
This work was supported in part by Natural Science Foundation of
Shandong Province, China under Grant No. ZR2015AQ006 and by
National Natural Science Foundation of China (Grants
No. 11521505, No. 11621131001, No. 11447009, No. 11505098, and No. 11765012).
R.~H. Li is partly supported by the plan of \emph{Young Creative Talents} under \emph{the Talent of the Prairie} project of the Inner Mongolia.

 \vspace{0.5 cm}

\appendix
\section{Results after momentum fraction integration}

\subsection{Tree level}
\subsubsection{Leading power in $v$}
\begin{eqnarray}
F^b_{(LO)}(Q^2) &=& if_PP_c \int_0^\infty db b \exp[-S_{\text{I}}
(x_0,x_0, b, Q, \mu)] H_0^{(1)}(x_0Qb) \phi^2_T(b),
\\
F^c_{(LO)}(Q^2) &=& if_PP_c \int_0^\infty db b \exp[-S_{\text{I}}
(x_0,x_0, b, Q, \mu)] H_0^{(1)}(\bar{x}_0Qb) \phi^2_T(b),
\\
A^b_{2,(LO)}(Q^2) &=& - \frac{f^\parallel_Vm_V P_c}{Q^2} \int_0^\infty db b
\exp[-S_{\text{I}} (x_0,x_0, b, Q, \mu)] H_0^{(1)}(x_0Qb)
\phi^2_T(b),
\\
A^c_{2,(LO)}(Q^2) &=& \frac{f^\parallel_Vm_V P_c}{Q^2} \int_0^\infty db b
\exp[-S_{\text{I}} (x_0,x_0, b, Q, \mu)] H_0^{(1)}(\bar{x}_0Qb)
\phi^2_T(b),
\end{eqnarray}
where $P_c=\pi^2f_PC_F\alpha_s(\mu)/N_c$, $\bar{x}_0=1-x_0$,
$S_{\text{I}} (x, y, b, Q,\mu) \equiv S (x, b, y, b, Q, \mu) $ and
$H^{(1)}_n(x)=J_n(x)+iY_n(x)$. We use the LQCD decay constant \cite{Chiu:2007km} $f_B = 0.489$~GeV.
The Sudakov factor is
\begin{align}
S(x,b_1,y,b_2,Q,\mu)=& s(x,b_1,Q)+s(\bar{x},b_1,Q) +2se(\mu,b_1)
\nonumber
\\
&+s(y,b_2,Q)+s(\bar{y},b_2,Q)+2se(\mu,b_2), \nonumber
\\
se(\mu,b)=& -\frac{1}{2\beta_1}\ln\frac{\ln(\mu^2/\Lambda^2)}
{\ln[1/(b^2\Lambda^2)]},
\end{align}

\begin{align}
s(\xi,b)~=~&\frac{A^{(1)}}{2\beta_1}\hat{q}\ln\Big(\frac{\hat{q}}{\hat{b}}\Big)-\frac{A^{(1)}}{2\beta_1}\Big(\hat{q}-\hat{b}\Big)
+\frac{A^{(2)}}{4\beta_1^2}\Big(\frac{\hat{q}}{\hat{b}}-1\Big)
-\frac{A^{(1)}\beta_2}{4\beta_1^3}\hat{q}\bigg[\frac{\ln(2\hat{b})+1}{\hat{b}}-\frac{\ln(2\hat{q})+1}{\hat{q}}\bigg] \nonumber\\
&-\bigg[\frac{A^{(2)}}{4\beta_1^2}-\frac{A^{(1)}}{4\beta_1}\ln\Big(\frac{e^{2\gamma_E-1}}{2}\Big)\bigg]\ln\Big(\frac{\hat{q}}{\hat{b}}\Big)+\frac{A^{(1)}\beta_2}{8\beta_1^3}\big[\ln^2(2\hat{q})-\ln^2(2\hat{b})\big],
\end{align}
where
\begin{align}
&\hat{q}\equiv\ln\Big[\frac{\xi
Q}{\sqrt{2}\Lambda_{\mathrm{QCD}}}\Big],~~~\hat{b}\equiv\ln\Big[\frac{1}{b\Lambda_{\mathrm{QCD}}}\Big],
\\
&A^{(1)}=\frac{4}{3},~~~A^{(2)}=\frac{67}{9}-\frac{\pi^2}{3}-\frac{10}{27}n_f+\frac{8}{3}\beta_1\ln\Big(\frac{e^{\gamma_E}}{2}\Big).
\end{align}

\subsubsection{NRQCD corrections}
{The $\delta^\prime$ term} $-\frac{ \langle
\textbf{q}^2\rangle}{(m_b+m_c)^2} \frac{2(1-2x_0)}{3x_0\bar
x_0}\delta'(x-x_0)$
\begin{eqnarray}
& F^{b,NR,I}_{(LO)}(Q^2) =  2i f_P P_c \frac{\langle \textbf{q}^2
\rangle}{(m_b+m_c)^2} \frac{2(1-2x_0)}{3 x_0 \bar{x}_0}\int^\infty_0db b
\text{exp}[-S_I(x_0,x_0,b,Q,\mu)] \nonumber
\\
& \times \bigg\{
\big[s_p(\bar{x}_0,b,Q)-s_p(x_0,b,Q)\big]H^{(1)}_0(x_0Qb) -
\frac{H^{(1)}_1(x_0Qb)}{2}Qb \bigg\}\phi^2_T(b),
\\
& F^{c,NR,I}_{(LO)}(Q^2) =  2i f_P P_c \frac{\langle \textbf{q}^2
\rangle}{(m_b+m_c)^2} \frac{2(1-2x_0)}{3 x_0 \bar{x}_0}\int^\infty_0db b
\text{exp}[-S_I(x_0,x_0,b,Q,\mu)] \nonumber
\\
& \times \bigg\{
\big[s_p(\bar{x}_0,b,Q)-s_p(x_0,b,Q)\big]H^{(1)}_0(\bar{x}_0Qb) +
\frac{H^{(1)}_1(\bar{x}_0Qb)}{2}Qb\bigg\} \phi^2_T(b),
\\
& A^{b,NR,I}_{2,(LO)}(Q^2) = i \frac{f^\parallel_Vm_V}{f_PQ^2}
F^{b,NR,I}_{(LO)}(Q^2),
\\
& A^{c,NR,I}_{2,(LO)}(Q^2) = -i \frac{f^\parallel_Vm_V}{f_PQ^2}
F^{c,NR,I}_{(LO)}(Q^2).
\end{eqnarray}
\begin{align}
& s_p(\xi,b,Q)=\frac{d s(\xi,b,Q)}{d \xi} \nonumber
\\
=&  \frac{1}{4\xi\beta_1}
\bigg\{A^{(1)}\Big(2\ln\frac{\hat{q}}{\hat{b}}
+\frac{1}{\hat{q}}\ln\frac{e^{2\gamma_E-1}}{2}\Big)
+\frac{A^{(1)}\beta_2}{\beta^2_1}
\Big[\frac{\ln(2\hat{q})+1}{\hat{q}}-\frac{\ln(2\hat{b})+1}{\hat{b}}\Big]
+\frac{A^{(2)}}{\beta_1}\Big(\frac{1}{\hat{b}}-\frac{1}{\hat{q}}\Big)
\bigg\},
\end{align}
\begin{align}
\frac{d s(\bar{\xi},b,Q)}{d\xi} =& - s_p(\bar{\xi},b,Q),
\\
\frac{d H^{(1)}_n(x)}{dx} =& \frac{nH^{(1)}_n(x)}{x}
-H^{(1)}_{n+1}(x).
\end{align}

{The $\delta^{\prime\prime}$ term} $\frac{ \langle
\textbf{q}^2\rangle}{6(m_b+m_c)^2}\delta''(x-x_0)$

\begin{eqnarray}
& F^{b,NR,II}_{(LO)}(Q^2) =  2i f_P P_c \frac{\langle \textbf{q}^2
\rangle}{6(m_b+m_c)^2} \int^\infty_0db b
\text{exp}[-S_I(x_0,x_0,b,Q,\mu)] \nonumber
\\
&\times\bigg\{ [s_p(\bar{x}_0)-s_p(x_0)] \Big\{
[s_p(\bar{x}_0)-s_p(x_0)]H^{(1)}_0(x_0Qb)
-\frac{Qb}{2}H^{(1)}_1(x_0Qb) \Big\} \nonumber
\\
&\!-\![s_{pp}(\bar{x}_0)+s_{pp}(x_0)]H^{(1)}_0(x_0Qb)
\!-\![s_p(\bar{x}_0)-s_p(x_0)]\frac{Qb}{2}H^{(1)}_1(x_0Qb)
\nonumber
\\
&+\!\frac{Q^2b^2}{4}H^{(1)}_2(x_0Qb)\bigg\} \phi^2_T(b) ,
\\
& F^{c,NR,II}_{(LO)}(Q^2) =  2i f_P P_c \frac{\langle \textbf{q}^2
\rangle}{6(m_b+m_c)^2} \int^\infty_0db b
\text{exp}[-S_I(x_0,x_0,b,Q,\mu)] \nonumber
\\
&\times\bigg\{ [s_p(\bar{x}_0)-s_p(x_0)] \Big\{
[s_p(\bar{x}_0)-s_p(x_0)]H^{(1)}_0(\bar{x}_0Qb)
+\frac{Qb}{2}H^{(1)}_1(\bar{x}_0Qb) \Big\} \nonumber
\\
&-[s_{pp}(\bar{x}_0)+s_{pp}(x_0)]H^{(1)}_0(\bar{x}_0Qb)
+[s_p(\bar{x}_0)-s_p(x_0)]\frac{Qb}{2} H^{(1)}_1(\bar{x}_0Qb)
\nonumber
\\
&+\frac{Q^2b^2}{4}H^{(1)}_2(\bar{x}_0Qb) \bigg\} \phi^2_T(b) ,
\\
& A^{b,NR,II}_{2,(LO)}(Q^2) = i \frac{f^\parallel_Vm_V}{f_PQ^2}
F^{b,NR,II}_{(LO)}(Q^2),
\label{eq.app1}
\\
& A^{c,NR,II}_{2,(LO)}(Q^2) = -i \frac{f^\parallel_Vm_V}{f_PQ^2}
F^{c,NR,II}_{(LO)}(Q^2),
\label{eq.app2}
\end{eqnarray}

\begin{align}
& s_{pp}(\xi,b,Q)=\frac{d s_p(\xi,b,Q)}{d \xi} \nonumber
\\
=&  \frac{1}{4\xi^2\beta_1} \bigg\{\frac{A^{(1)}}{\hat{q}}
\Big[2-\frac{\ln(e^{2\gamma_E-1}/2)}{\hat{q}}\Big]
-\frac{A^{(1)}\beta_2}{\beta^2_1}\frac{\ln(2\hat{q})}{\hat{q}^2}
+\frac{A^{(2)}}{\beta_1\hat{q}^2} \nonumber
\\
-&\bigg[A^{(1)}\Big(2\ln\frac{\hat{q}}{\hat{b}}
+\frac{1}{\hat{q}}\ln\frac{e^{2\gamma_E-1}}{2}\Big)
+\frac{A^{(1)}\beta_2}{\beta^2_1}
\Big[\frac{\ln(2\hat{q})+1}{\hat{q}}-\frac{\ln(2\hat{b})+1}{\hat{b}}\Big]
+\frac{A^{(2)}}{\beta_1}\Big(\frac{1}{\hat{b}}-\frac{1}{\hat{q}}\Big)\bigg]
\bigg\}.
\end{align}

\subsection{One-loop level}

{The NLO correction from the hard kernel is:}
\begin{eqnarray}
F^{b,I}_{(NLO)}(Q^2) =  & if_P P_c\frac{\alpha_s(\mu)C_F}{4\pi}
\int_0^\infty db b \, \exp[-S_{\text{I}} (x_0, x_0, b, Q, \mu)]
\phi^2_T(b) \nonumber
\\
&\times \left[ \,\widetilde{h}(x_0, x_0, b, Q, \mu) \,
H_0^{(1)}(x_0Qb) +\frac{\ln^2(x_0Qb)}{3}H^{(1)}_0(x_0Qb) \right] ,
\\
F^{c,I}_{(NLO)}(Q^2) =  & if_P P_c\frac{\alpha_s(\mu)C_F}{4\pi}
\int_0^\infty db b \, \exp[-S_{\text{I}} (x_0, x_0, b, Q, \mu)]
\phi^2_T(b) \nonumber
\\
&\times \left[ \,\widetilde{h}(\bar{x}_0, \bar{x}_0, b, Q, \mu) \,
H_0^{(1)}(\bar{x}_0Qb)
+\frac{\ln^2(\bar{x}_0Qb)}{3}H^{(1)}_0(\bar{x}_0Qb) \right].
\end{eqnarray}

We add the one-loop corrections from the two wave functions
together. We rewrite the wave function as :
\begin{align}
\hat{\phi}^{(1,0)}_P(x) = & \frac{\alpha_sC_F}{4\pi}
[\hat{\phi}^I(x)+\hat{\phi}^{II}(x)],
\\
\hat{\phi}^I(x) = &2\left[ \left(
\ln{\frac{\mu^2}{(m_b+m_c)^2(x_0-x)^2}}-1 \right) \frac{x_0+\bar
x}{x_0-x}\frac{x}{x_0}\theta(x_0-x) \right]_+ \nonumber
\\
+2&\left[ \left( \ln{\frac{\mu^2}{(m_b+m_c)^2(x_0-x)^2}}-1 \right)
\frac{\bar{x}_0+x}{\bar{x}_0-\bar{x}}\frac{\bar{x}}{\bar{x}_0}\theta(x-x_0)\right]_+
+\left[ \frac{4x\bar x}{(x_0-x)^2}\right]_{++},
\\
\hat{\phi}^{II}(x) = & \left[ 4x_0\bar x_0 \ln{\frac{x_0}{\bar
x_0}}+2(2x_0-1) \right]\delta^\prime(x-x_0).
\end{align}

{The NLO corrections from $\hat{\phi}^{I}$ are:}
\begin{eqnarray}
F^{b,II}_{(NLO)}(Q^2) =& 2if_PP_c\frac{\alpha_sC_F}{4\pi}
\int_0^1\!\!dx \int_0^\infty \!\!db b \Big\{\exp[-S_{\text{I}}
(x,x_0, b, Q, \mu)]H_0^{(1)}(\sqrt{x x_0}Qb) \nonumber
\\
& -\exp[-S_{\text{I}} (x_0,x_0, b, Q, \mu)] H_0^{(1)}(x_0Qb)\Big\}
\Psi(x) \phi^2_T({x},b) \nonumber
\\
&-2if_PP_c\frac{\alpha_sC_F}{4\pi} \int_0^1\!\!dx \int_0^\infty
\!\!db b \frac{4x\bar x}{(x-x_0)}
\text{exp}[-S_I(x_0,x_0,b,Q,\mu)] \nonumber
\\
& \times \bigg\{
\big[s_p(\bar{x}_0,b,Q)-s_p(x_0,b,Q)\big]H^{(1)}_0(x_0Qb) -
\frac{H^{(1)}_1(x_0Qb)}{2}Qb \bigg\} \phi^2_T(b),
\\
F^{c,II}_{(NLO)}(Q^2) =& 2if_PP_c\frac{\alpha_sC_F}{4\pi}
\int_0^1\!\!dx \int_0^\infty \!\!db b \Big\{\exp[-S_{\text{I}}
(x,x_0, b, Q, \mu)]H_0^{(1)}(\sqrt{\bar{x}\bar{x}_0}Qb) \nonumber
\\
& -\exp[-S_{\text{I}} (x_0,x_0, b, Q, \mu)]
H_0^{(1)}(\bar{x}_0Qb)\Big\} \Psi(x) \phi^2_T({x},b) \nonumber
\\
&-2if_PP_c\frac{\alpha_sC_F}{4\pi} \int_0^1\!\!dx \int_0^\infty
\!\!db b \frac{4x\bar x}{(x-x_0)}
\text{exp}[-S_I(x_0,x_0,b,Q,\mu)] \nonumber
\\
& \times \bigg\{
\big[s_p(\bar{x}_0,b,Q)-s_p(x_0,b,Q)\big]H^{(1)}_0(\bar{x}_0Qb) +
\frac{H^{(1)}_1(\bar{x}_0Qb)}{2}Qb \bigg\}\phi^2_T(b),
\end{eqnarray}
\begin{align}
\Psi(x) = &2\left( \ln{\frac{\mu^2}{(m_b+m_c)^2(x_0-x)^2}}-1
\right) \frac{x_0+\bar x}{x_0-x}\frac{x}{x_0}\theta(x_0-x)
\nonumber
\\
&+2\left( \ln{\frac{\mu^2}{(m_b+m_c)^2(x_0-x)^2}}-1 \right)
\frac{\bar{x}_0+x}{x-x_0}\frac{\bar{x}}{\bar{x}_0}\theta(x-x_0)
+\frac{4x\bar x}{(x_0-x)^2}.
\end{align}

{The NLO corrections from $\hat{\phi}^{II}$ are:}
\begin{eqnarray}
F^{b,III}_{(NLO)}(Q^2)= & \frac{\alpha_sC_F}{4\pi} \left[
4x_0\bar{x}_0 \ln{\frac{x_0}{\bar x_0}}+2(2x_0-1)\right]/ \left[
-\frac{ \langle \textbf{q}^2\rangle}{(m_b+m_c)^2}
\frac{2(1-2x_0)}{3x_0\bar x_0} \right]F^{b,NR,I}_{(LO)}(Q^2),
\\
F^{c,III}_{(NLO)}(Q^2)= & \frac{\alpha_sC_F}{4\pi} \left[
4x_0\bar{x}_0 \ln{\frac{x_0}{\bar x_0}}+2(2x_0-1)\right]/ \left[
-\frac{ \langle \textbf{q}^2\rangle}{(m_b+m_c)^2}
\frac{2(1-2x_0)}{3x_0\bar x_0} \right]F^{c,NR,I}_{(LO)}(Q^2).
\end{eqnarray}

{The NLO corrections from the decay constant and Sudakov factor are:}
\begin{align}
F^{b,IV}_{(NLO)}(Q^2) =& i(2f^{NLO}_P)f_PP_c \int_0^\infty db b
\exp[-S_{\text{I}} (x_0,x_0, b, Q, \mu)]
H_0^{(1)}(\bar{x}_0Qb)\phi^2_T(b) \nonumber
\\
-if_PP_c  & \int_0^\infty db b \exp[-S_{\text{I}} (x_0,x_0, b,
Q, \mu)] S^{NLO}_{\text{I}} (x_0,x_0, b,
Q, \mu) H_0^{(1)}(x_0Qb) \phi^2_T(b),
\\
F^{c,IV}_{(NLO)}(Q^2) =& i(2f^{NLO}_P)f_PP_c \int_0^\infty db b
\exp[-S_{\text{I}} (x_0,x_0, b, Q,
\mu)]H_0^{(1)}(\bar{x}_0Qb)\phi^2_T(b) \nonumber
\\
-if_PP_c & \int_0^\infty db b \exp[-S_{\text{I}} (x_0,x_0, b,
Q, \mu)] S^{NLO}_{\text{I}} (x_0,x_0, b,
Q, \mu) H_0^{(1)}(\bar{x}_0Qb) \phi^2_T(b),
\end{align}
where $S^{NLO}_{\text{I}} (x, y, b, Q,\mu) \equiv S^{NLO} (x, b,
y, b, Q, \mu)$.

\begin{align}
&f^{NLO}_P =  \frac{\alpha_s(m_{B_c})C_F}{4\pi} \left(
3\ln\frac{m_{B_c}}{m_c}-4 \right), \nonumber
\\
&S^{NLO}(x,b_1,y,b_2,Q,\mu)=
s_{NLO}(x,b_1,Q)+s_{NLO}(\bar{x},b_1,Q) +2se_{NLO}(\mu,b_1)
\nonumber
\\
&\qquad +s_{NLO}(y,b_2,Q)+s_{NLO}(\bar{y},b_2,Q)
+2se_{NLO}(\mu,b_2), \nonumber
\\
&se_{NLO}(\mu,b)=\frac{\beta_2}{2\beta^3_1} \bigg\{
\frac{\ln\ln[1/(b^2\Lambda^2)]+1}{\ln[1/(b^2\Lambda^2)]}
-\frac{\ln\ln(\mu^2/\Lambda^2)+1}{\ln(\mu^2/\Lambda^2)} \bigg\},
\end{align}
\begin{align}
s_{NLO}(\xi,b)= &\frac{A^{(1)}\beta_{2}}{8\beta_{1}^{3}}
\ln\left(\frac{e^{2\gamma-1}}{2}\right)\left[
\frac{\ln(2\hat{q})+1}{\hat{q}}-\frac{\ln(2\hat{b})+1}{\hat{b}}\right]
-\frac{A^{(1)}\beta_{2}}{16\beta_{1}^{4}}
\frac{\hat{q}-\hat{b}}{\hat{b}^2}\left[2\ln(2\hat{b})+1\right]
\nonumber \\
&-\frac{A^{(1)}\beta_{2}}{16\beta_{1}^{4}}\left[
\frac{2\ln(2\hat{q})+3}{\hat{q}}-\frac{2\ln(2\hat{b})+3}{\hat{b}}\right]
+\frac{A^{(2)}\beta_{2}^2}{432\beta_{1}^{6}}
\frac{\hat{q}-\hat{b}}{\hat{b}^3}
\left[9\ln^2(2\hat{b})+6\ln(2\hat{b})+2\right]
\nonumber \\
&+\frac{A^{(2)}\beta_{2}^2}{1728\beta_{1}^{6}}\left[
\frac{18\ln^2(2\hat{q})+30\ln(2\hat{q})+19}{\hat{q}^2}
-\frac{18\ln^2(2\hat{b})+30\ln(2\hat{b})+19}{\hat{b}^2}\right],
\end{align}

\begin{align}
F^b(Q^2)= & F^{b}_{(LO)}+F^{b,NR,I}_{(LO)}+F^{b,NR,II}_{(LO)}+
F^{b,I}_{(NLO)}+F^{b,II}_{(NLO)}+F^{b,III}_{(NLO)}+F^{b,IV}_{(NLO)},
\\
F^c(Q^2)= & F^{c}_{(LO)}+F^{c,NR,I}_{(LO)}+F^{c,NR,II}_{(LO)}+
F^{c,I}_{(NLO)}+F^{c,II}_{(NLO)}+F^{c,III}_{(NLO)}+F^{c,IV}_{(NLO)},
\\
A^b_2(Q^2)= &
A^{b}_{2,(LO)}+A^{b,NR,I}_{2,(LO)}+A^{b,NR,II}_{2,(LO)},
\\
A^c_2(Q^2)= &
A^{c}_{2,(LO)}+A^{c,NR,I}_{2,(LO)}+A^{c,NR,II}_{2,(LO)}.
\end{align}

\clearpage

\end{CJK*}
\end{document}